\documentclass{aa}
\usepackage{graphicx,natbib,amsmath,amssymb,multirow}
\usepackage[T1]{fontenc}
\usepackage{mathptmx}
\usepackage[colorlinks=true,citecolor=blue,urlcolor=blue]{hyperref}
\bibpunct{(}{)}{;}{a}{}{,}
\newcommand{\kms}{\ifmmode{~{\rm km\,s}^{-1}}\else{~km~s$^{-1}$~}\fi}
\newcommand{\msun}{\ifmmode{{\rm M}_{\odot}}\else{${\rm M}_{\odot}$~}\fi}
\newcommand{\msunit}{\ifmmode{M_{\odot}}\else{$M_{\odot}$~}\fi}
\newcommand{\lsun}{\ifmmode{{\rm L}_{\odot}}\else{${\rm L}_{\odot}$}\fi}
\newcommand{\cbone}{\rlap{~{\small$\mathrm{\mathsf{1}}$}}{$\bigcirc$}}
\newcommand{\cbtwo}{\rlap{~{\small$\mathrm{\mathsf{2}}$}}{$\bigcirc$}}
\newcommand{\cbones}{\rlap{~{\scriptsize$\mathrm{\mathsf{1}}$}}{$\bigcirc$}}
\newcommand{\cbtwos}{\rlap{~{\scriptsize$\mathrm{\mathsf{2}}$}}{$\bigcirc$}}

\defcitealias{vaytet2012}{Paper~I}
\defcitealias{masunaga2000}{MI00}
\defcitealias{tomida2013}{Tea13}
\defcitealias{saumon1995}{SCvH95}
\defcitealias{stamatellos2007}{SWBG07}

\begin{document}

\title{On the role of the $\text{H}_{2}$ ortho:para ratio in gravitational\\ collapse during star formation}
\titlerunning{On the role of the $\text{H}_{2}$ ortho:para ratio in star formation}

\author{Neil Vaytet$^{1}$, Kengo Tomida$^{2}$, Gilles Chabrier$^{1,3}$}
\authorrunning{Vaytet et al.}

\institute{$^{1}$ \'{E}cole Normale Sup\'{e}rieure de Lyon, CRAL, UMR CNRS 5574, Universit\'{e} Lyon I, 46 All\'{e}e d'Italie, 69364 Lyon Cedex 07, France\\
           $^{2}$ Department of Astrophysical Sciences, Princeton University, Princeton, NJ 08544, USA\\
           $^{3}$ School of Physics, University of Exeter, Exeter, EX4 4QL, UK
          }

\date{Received / Accepted}

\offprints{neil.vaytet@ens-lyon.fr} 

\abstract{Hydrogen molecules ($\text{H}_{2}$) come in two forms in the interstellar medium, ortho- and para-hydrogen, corresponding to
the two different spin configurations of the two hydrogen atoms. The relative abundances of the two flavours in the interstellar medium
are still very uncertain, and this abundance ratio has a significant impact on the thermal properties of the gas. In the context of
star formation, theoretical studies have recently adopted two different strategies when considering the ortho:para ratio (OPR) of 
$\text{H}_{2}$ molecules; the first considers the OPR to be frozen at 3:1 while the second assumes that the species are in thermal
equilibrium at all temperatures.}
{As the OPR potentially affects the protostellar cores which form as a result of the gravitational collapse of a dense molecular cloud,
the aim of this paper is to quantify precisely what role the choice of OPR plays in the properties and evolution of the cores.}
{We used two different ideal gas equations of state for a hydrogen and helium mix in a radiation hydrodynamics code to simulate the
collapse of a dense cloud and the formation of the first and second Larson cores; the first equation of state uses a fixed OPR of
3:1 while the second assumes thermal equilibrium.}
{The OPR was found to markedly impact the evolution of the first core. Simulations using an equilibrium ratio collapse faster at early
times and show noticeable oscillations around hydrostatic equilibrium, to the point where the core expands for a short time
right after its formation before resuming its contraction. In the case of a fixed 3:1 OPR, the core's evolution is a lot smoother.
The OPR was however found to have little impact on the size, mass and radius of the two Larson cores.}
{It is not clear from observational or theoretical studies of OPR in molecular clouds which OPR should be used in the context of star
formation. Our simulations show that if one is solely interested in the final properties of the cores when they are formed, it does not
matter which OPR is used. On the other hand, if one's focus lies primarily in the evolution of the first core, the choice of OPR
becomes important.}


\maketitle

\section{Introduction}\label{sec:intro}

Hydrogen molecules ($\text{H}_{2}$) come in two forms in the interstellar medium, corresponding to the two different spin configurations
of the two hydrogen atoms. The first one, often called ortho-hydrogen, is a triplet state where the two proton spins are aligned in
parallel fashion, while the second, known as para-hydrogen, is a singlet state with the two proton spins aligned in an anti-parallel
manner. The relative abundances of the two flavours in the interstellar medium are still very uncertain. When the molecules form on the
surface of dust grains, the ortho:para ratio (OPR) is believed to be 3:1, reflecting the statistical weight of each variety according to
their spin degeneracies \citep[see for instance][]{dyson1997,duley1984,duley1993,takahashi2001,habart2005,gavilan2012}.

The transition back to ortho-para equilibrium is known to be a lengthy process, unless a catalyst is present in the medium (ortho-para
conversion may occur through proton exchange reactions between $\text{H}_{2}$ and other species), as there are no radiative transitions
between ortho- and para-hydrogen \citep{raich1964,souers1986,habart2005}. This implies that, at low temperatures ($T \lesssim 300$ K), the
population distribution of the two $\text{H}_{2}$ monomers is not the equilibrium value. Observations indeed suggest that the real
abundance ratio in molecular clouds and star forming regions is far from the thermal equilibrium value (see \citealt{pagani2011} and
\citealt{dislaire2012} for two recent examples), even though large discrepancies (due to observational difficulties) between the studies
remain. \citet{flower1984} have also shown through theoretical calculations that, under typical molecular cloud conditions ($n \sim 100
- 1000~\text{cm}^{-3}$), it takes al least 1 Myr for the gas to reach ortho:para equilibrium \citep[see also][]{flower2006}. The
conversions between para- and ortho-hydrogen states and their relative abundances under astrophysical conditions have been studied by many
authors; in molecular clouds \citep{osterbrock1962,dalgarno1973,rodriguez2000,crabtree2011}, jovian planets
\citep{decampli1978,massie1982,carlson1992}, protostellar systems \citep{boley2007,pagani2009}, jets
\citep{smith1997,neufeld1998,neufeld2006}, nebulae \citep{takayanagi1987,hoban1991} and even other galaxies \citep{harrison1998}.

The OPR has a significant impact on the thermodynamic properties of the gas, primarily on the heat capacity of the gas (see
section~\ref{sec:eos}), and the choice of OPR has potential important implications for simulations of star formation from the collapse
of dense molecular cloud core. Recent studies have used both non-equilibrium \citep{stamatellos2007,tomida2013} and equilibrium
\citep{masunaga2000,vaytet2013} treatments of the hydrogen isomers, reporting different thermal evolutions during the formation and
subsequent contraction and mass growth of the first Larson core. A comparison of the different studies was carried out by
\citet{vaytet2013}, and they concluded that although the OPR appeared not to be the dominant source of discrepancies between the
different simulations, the only way to be sure would be to run two simulations with the same code (using the exact same method
for radiative transfer), adopting equilibrium and non-equilibrium OPRs in the two different cases.

This is precisely the aim of this paper; we will first describe the simulation setup, including the numerical method and the different
equations of state (EOS) used to model the gas thermodynamics, then we discuss the results of the simulations and the impact of the
choice of OPR on the properties of the first and second Larson cores.

\section{Description of the simulations}\label{sec:simulations}

\subsection{Numerical method and setup}\label{sec:setup}

The code used to solve the multigroup RHD equations is a one-dimensional fully implicit Godunov Lagrangian code described in \citet{vaytet2013}.
It uses the $M_{1}$ closure to model the radiative transfer \citep{levermore1984,dubroca1999} and the grid comprises 2000 cells logarithmically
spaced in the radial direction. The interstellar dust and gas opacities used were also identical to that of \citet{vaytet2013}.

The initial setup for the dense core collapse was taken from \citet{vaytet2013}. A uniform density sphere of mass $M_{0} = 1~\msun$,
temperature $T_{0} = 10$ K ($c_{s0} = 0.187 \kms$) and radius $R_{0} = 10^{4}$ AU collapses under its own gravity. The cloud's free-fall time
is $t_{\text{ff}} \simeq 177$ Kyr. The radiation temperature is in equilibrium with the gas temperature and the radiative flux is set to zero
everywhere. The boundary conditions are reflexive at the centre of the grid ($r = 0$) and all the variables at the outer edge of the sphere are
fixed to their initial values. The equations of radiative transfer were integrated over all frequencies (grey approximation) since including
frequency dependence only yields small differences for a much increased computational cost \citep{vaytet2012,vaytet2013}.

\subsection{Gas equations of state}\label{sec:eos}

To assess the effects of the ratio of ortho- to para-hydrogen on the collapse of a molecular cloud core, we used two different EOSs. Both model
the behaviour of an ideal mixture of hydrogen and helium considering the species $\text{H}_{2}$, H, $\text{H}^{+}$, He, $\text{He}^{+}$, and
$\text{He}^{2+}$. The He mass concentration was 0.27 and the full details on the computations of the different partition functions and
thermodynamic quantities can be found in \citet[Appendix A]{tomida2013}. The first EOS table (A) uses a fixed ortho:para ratio of 3:1 while
the second (B) assumes thermal equilibrium at all temperatures. The partition function of rotational transitions of molecular hydrogen for the
3:1 OPR is
\begin{equation}
Z_{\mathrm{rot}}^{\mathrm{ne}} = \left(Z_{\mathrm{rot}}^{\mathrm{even}}\right)^{\frac{1}{4}}\left[3 Z_{\mathrm{rot}}^{\mathrm{odd}} \exp\left(\frac{2\theta_{\mathrm{rot}}}{T}\right)\right]^{\frac{3}{4}} ~,
\end{equation}
(see \citealt{wannier1966}; \citealt{schwabl2006} for a derivation, and \citealt{boley2007} for the origin of the normalisation factor on the odd part)
while the one for the equilibrium model is
\begin{equation}
Z_{\mathrm{rot}}^{\mathrm{eq}} = Z_{\mathrm{rot}}^{\mathrm{even}}+3Z_{\mathrm{rot}}^{\mathrm{odd}} ~,
\end{equation}
where
\begin{alignat}{1}
Z_{\mathrm{rot}}^{\mathrm{even}} & = \sum_{j=0,2,4...} (2 j + 1) \exp \left[ -\frac{j(j+1)\theta_{\mathrm{rot}}}{T}\right] ~,\\
Z_{\mathrm{rot}}^{\mathrm{odd}}  & = \sum_{j=1,3,5...} (2 j + 1) \exp \left[ -\frac{j(j+1)\theta_{\mathrm{rot}}}{T}\right] ~,
\end{alignat}
and $\theta_{\mathrm{rot}} = 85.32$ K is the rotational excitation temperature.

\begin{figure}
\centering
\includegraphics[scale=0.41]{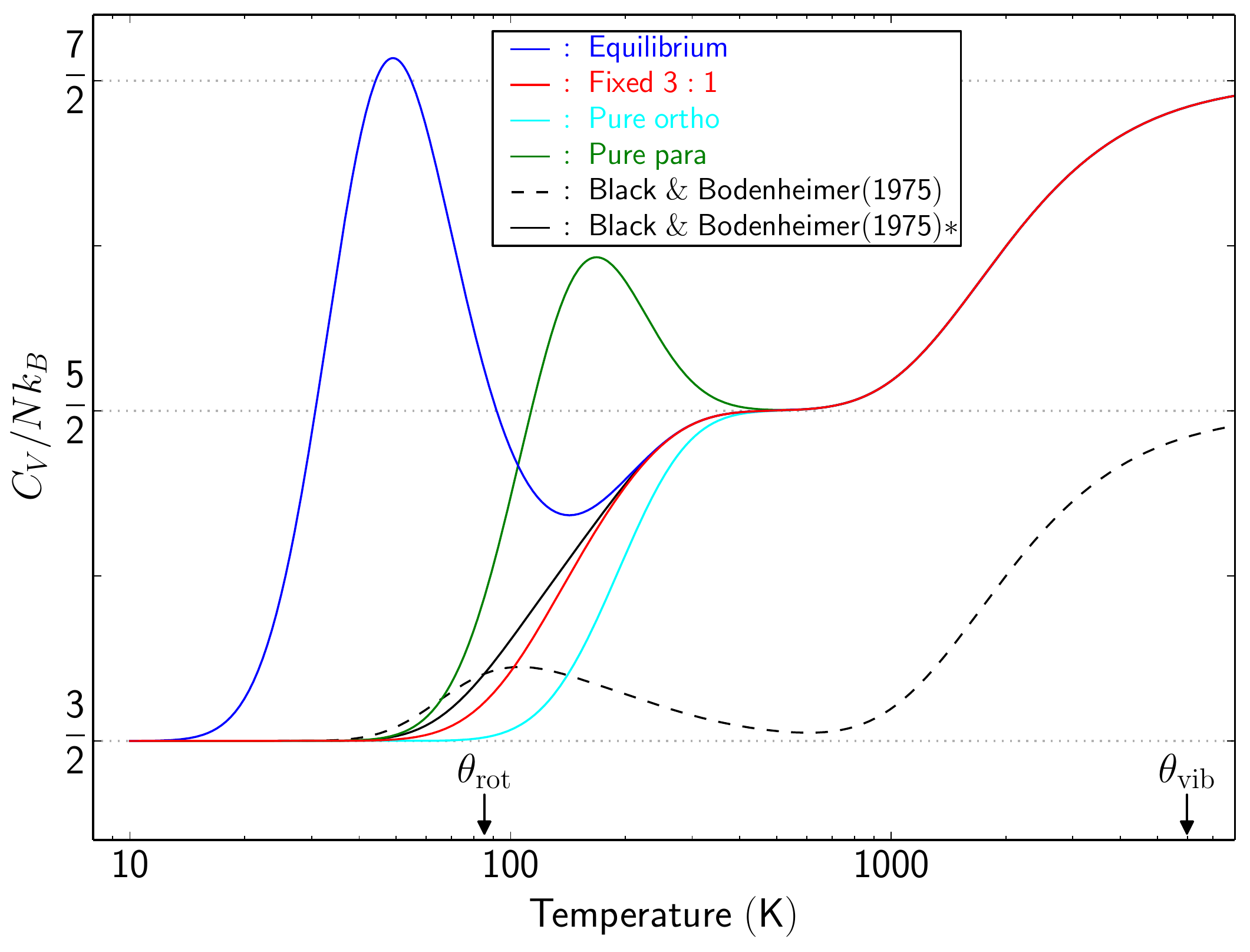}
\caption{Heat capacity for molecular hydrogen using different ortho:para ratios (see legend for details). $\theta_{\mathrm{rot}} = 85.32$ K
and $\theta_{\mathrm{vib}} = 5984.48$ K are the rotational and vibrational excitation temperatures, respectively. The black dashed curve
represents the raw model from \citet{black1975} while the black solid curve is the modified \citet{black1975} formula (see text).}
\label{fig:cv}
\end{figure}

The choice of ortho:para ratio has a direct impact on the heat capacity of the gas. In Fig.~\ref{fig:cv} are displayed the heat capacity
of $\text{H}_{2}$ at constant volume $C_{V}$ as a function of temperature for different treatments of ortho- and para-hydrogen (see for
instance \citealt{balian2007} for the computation of $C_{V}$). The blue curve represents the heat capacity assuming thermal equilibrium at
all temperatures, while the red curve is obtained assuming a constant 3:1 OPR. For completeness, we have also included the green and cyan
curves representing the pure para and pure ortho cases, respectively. Finally, we have also plotted the model by \citet{black1975} which
presents two peculiarities. The first is that if we use the formula as written in their equation (11), we get the dashed curve, which is
clearly wrong and also inconsistent with their Fig.~1. In order to recover the correct shape for the $C_{V}$ curve, it was necessary to
change the rotational contribution from $(\theta_{\text{rot}} / T_{m} )^{2} f(T_{m})$ to simply $f(T_{m})$; this yields the
modified (*) black solid line in Fig.~\ref{fig:cv}. The second point is that they explicitely write that their equation (13) holds for the
case where the ortho and para states are in equilibrium at all temperatures, yet their curve resembles a 3:1 ratio much more than the
equilibrium model in Fig.~\ref{fig:cv}. It should be also noted that \citet{boley2007} pointed out that \citet{black1975} used an
inadequate formula to calculate the internal energy, $e = C_{V}T$, while the correct definition is $C_{V} = de/dT$. The first relation is
valid only when $C_{V}$ does not depend on the temperature, and thus results in erroneous thermodynamic behaviour.

Finally, we also used for comparison purposes a third EOS (C) by \citet{saumon1995} (and its extension to low densities; see
\citealt{vaytet2013}) which models the properties of the same mixture of gas but also includes non-ideal effects at high densities. It
assumes an equilibrium ratio of ortho:para hydrogen and should behave very similarly to EOS B, at least at low-to-moderate densities.

\section{Results}\label{sec:results}

\subsection{Thermal evolution}\label{sec:thermal_evolution}

\begin{figure*}
\centering
\includegraphics[scale=0.67]{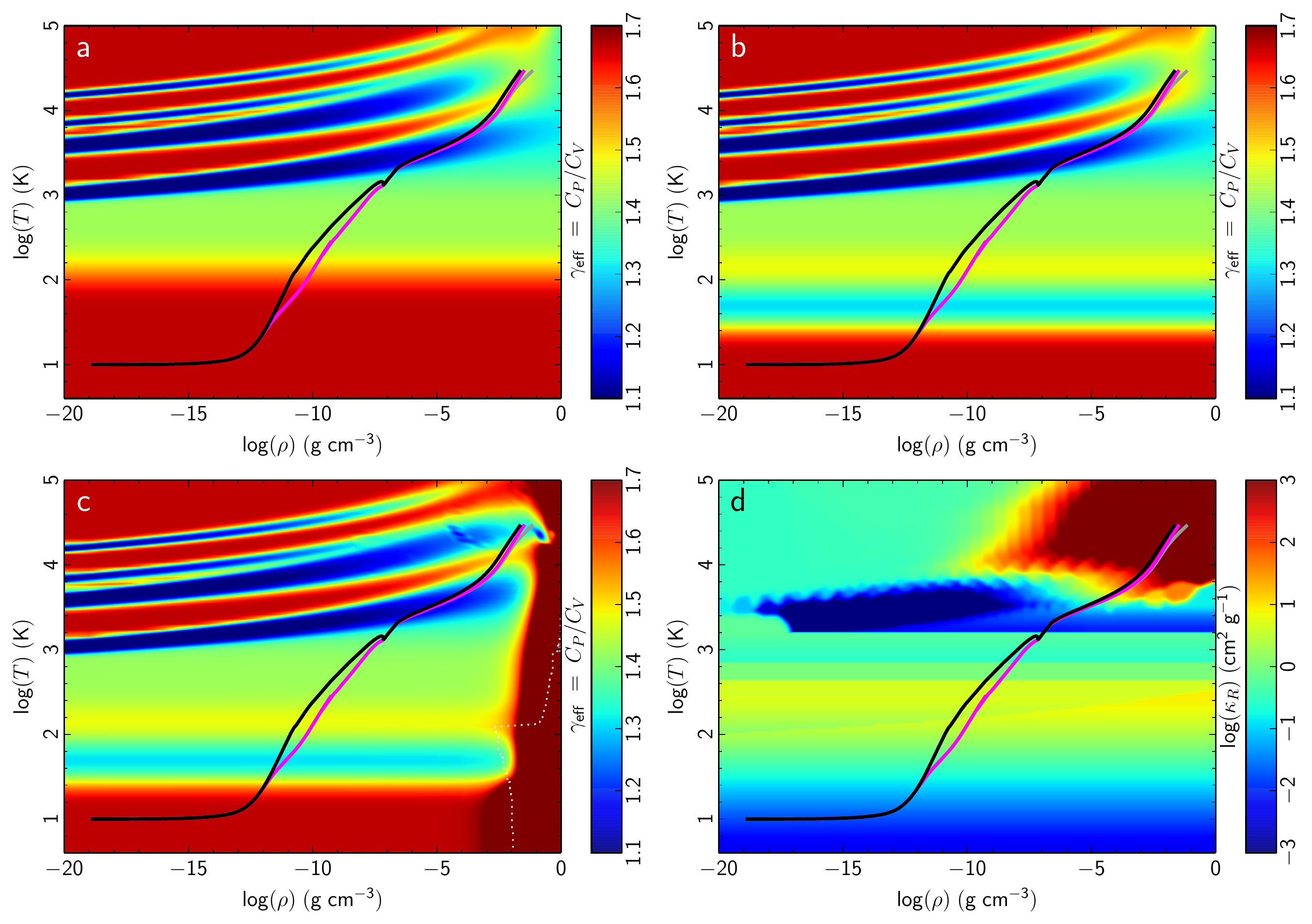}
\caption{Thermal evolution using EOS A (black; 3:1 OPR), EOS B (magenta; equilibrium OPR) and EOS C (grey; \citealt{saumon1995}). The
effective ratio of specific heats $\gamma_{\text{eff}}$ is displayed in color in the background for EOSs A, B, and C in panels (a), (b),
and (c), respectively. The region in the lower right corner of panel (c) delineated by the dotted white line indicates the area of the
($\rho,T$) space where the values in the EOS table cannot be trusted. In panel (d), the Rosseland mean opacity displayed in the
background.}
\label{fig:rhoT_all}
\end{figure*}

Three simulations were carried out; run 1 was performed using the fixed 3:1 EOS A, run 2 using the equilibrium EOS B, and run 3 using the
\citet{saumon1995} EOS C (see Table~\ref{tab:params}). The simulations were stopped when the temperature at the centre of the grid reached
30,000 K. Figure~\ref{fig:rhoT_all} shows the thermal evolution of the gas at the centre of the grid for all three runs (solid lines). The
effective ratio of specific heats $\gamma_{\text{eff}}$ of EOSs A, B, and C are displayed in color in the background of panels (a), (b),
and (c), respectively. In panel (d), the color background is used to represent the Rosseland mean opacity $\kappa_{R}$. We defined
$\gamma_{\text{eff}} = \rho c_{S}^{2} / P$, where $\rho$, $c_{S}$, and $P$ are the gas density, sound speed, and pressure, respectively.

\begin{table*}
\centering
\caption[Simulation parameters]{Simulation parameters. Columns 2 and 3 list the type of EOS used and the corresponding OPR. Columns 4 and 5
indicate the initial cloud radius and free-fall time. Columns 6 to 11 report the first and second cores' masses ($M_{1}$, $M_{2}$), radii
($R_{1}$, $R_{2}$) and entropies ($S_{1}$, $S_{2}$). The entropies are measured at the centre of the cores, and the first core entropy
corresponds to the time when the central density reaches $10^{-8}~\text{g~cm}^{-3}$.}
\begin{tabular}{ccccccccccc}
\hline
Run    & EOS & OPR & $R_{\text{init}}$                  & $t_{\text{ff}}$      & $M_{1}$             & $S_{1}$              & $R_{1}$ & $M_{2}$             & $R_{2}$            & $S_{2}$             \\
number & ~   & ~   & (AU)                               & (Kyr)                & (\msun)             & (erg/K/g)            & (AU)    & (\msun)             & (AU)               & (erg/K/g)           \\
\hline
 1     & A   & 3:1 & \multirow{3}{*}{$10^{4}$}          & \multirow{3}{*}{177} & $4.61\times10^{-2}$ & $1.03\times10^{10}$  & 26.7    & $1.76\times10^{-3}$ & $4.3\times10^{-3}$ & $1.01\times10^{10}$ \\
 2     & B   & Equ &                                    &                      & $4.28\times10^{-2}$ & $9.85\times10^{9}$   & 24.8    & $1.41\times10^{-3}$ & $3.4\times10^{-3}$ & $9.79\times10^{9}$  \\
 3     & C   & Equ &                                    &                      & $4.26\times10^{-2}$ & $9.79\times10^{9}$   & 24.8    & $1.39\times10^{-3}$ & $3.2\times10^{-3}$ & $9.81\times10^{9}$  \\
\hline
 4     & A   & 3:1 & \multirow{3}{*}{$5 \times 10^{3}$} & \multirow{3}{*}{62}  & $2.65\times10^{-2}$ & $1.06\times10^{10}$  &  5.9    & $2.13\times10^{-3}$ & $5.0\times10^{-3}$ & $1.05\times10^{10}$ \\
 5     & B   & Equ &                                    &                      & $2.42\times10^{-2}$ & $9.99\times10^{9}$   &  6.4    & $1.52\times10^{-3}$ & $3.6\times10^{-3}$ & $1.00\times10^{10}$ \\
 6     & C   & Equ &                                    &                      & $2.36\times10^{-2}$ & $9.93\times10^{9}$   &  6.3    & $1.46\times10^{-3}$ & $3.3\times10^{-3}$ & $1.00\times10^{10}$ \\
\hline
\end{tabular}
\label{tab:params}
\end{table*}

\begin{figure*}
\centering
\includegraphics[scale=0.68]{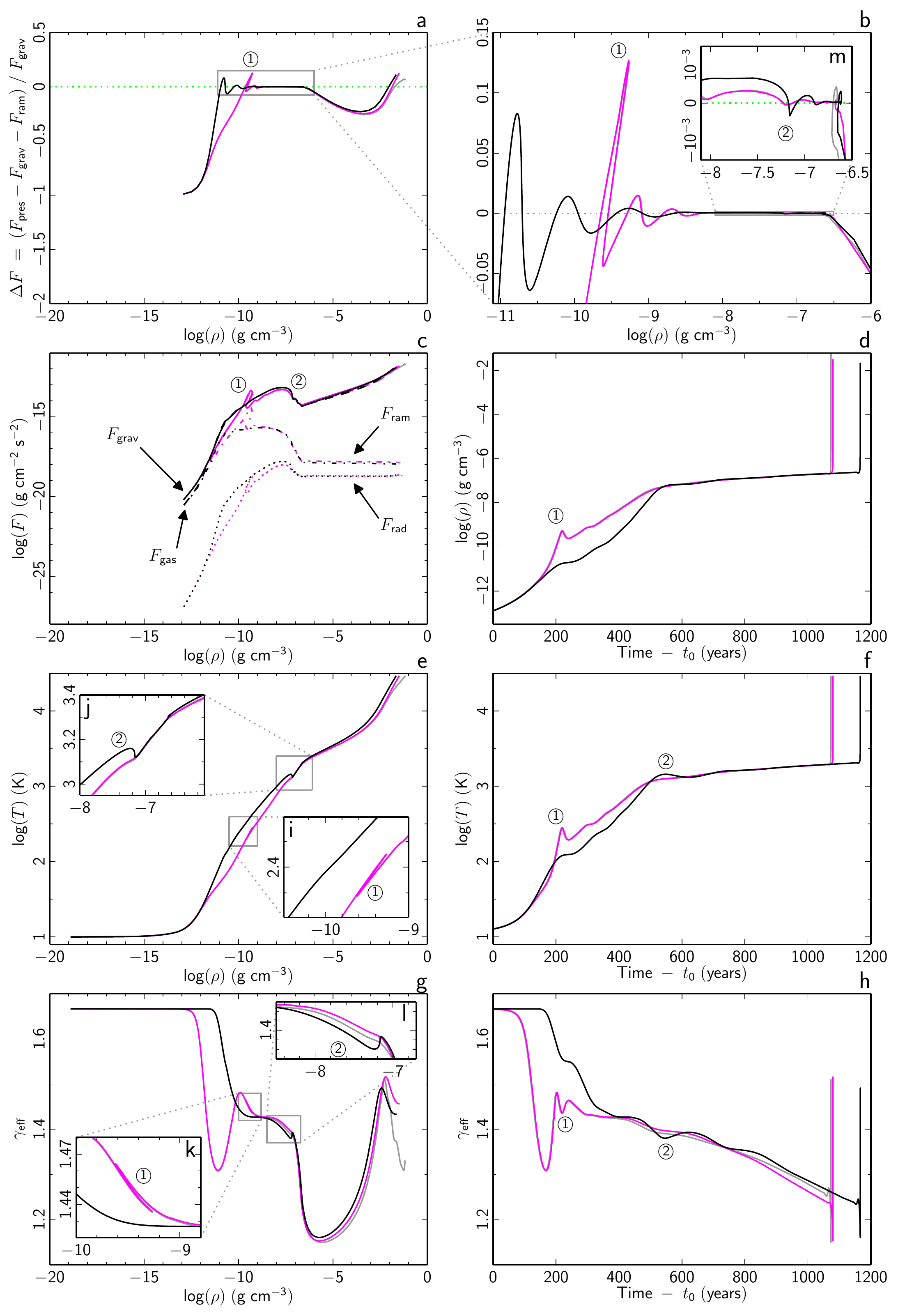}
\caption{Evolutions of the collapsing systems as a function of central density and time using EOS A (black), EOS B (magenta), and EOS C
(grey). (a,b) Normalised force differential $\Delta F$ between total pressure $F_{\text{pres}}$, gravity $F_{\text{grav}}$, and ram
pressure $F_{\text{ram}}$ (integrated over the volume of the first core) as a function of central density. The green dotted
$\Delta F = 0$ line represents hydrostatic equilibrium. (c) The separate force components from gravity $F_{\text{grav}}$ (solid), gas
pressure $F_{\text{gas}}$ (dashed), radiation pressure $F_{\text{rad}}$ (dotted), and ram pressure $F_{\text{ram}}$ (dot-dash) as a
function of central density. (d) Central density as a function of time. (e) Central temperature as a function of central density. (f)
Central temperature as a function of time. (g) Adiabatic index along the thermal tracks from panel (e). (h) Adiabatic index at the
centre of the grid as a function of time. The time of first core formation $t_{\text{0}}$ is assumed to be when the central density
reaches $10^{-13}~\text{g~cm}^{-3}$. In all panels, two thermal bounces are labeled {\cbones} and {\cbtwos} (see text).}
\label{fig:forces}
\end{figure*}

It can be clearly seen that using a different treatment of ortho- and para-hydrogen leads to significant differences in the thermal
evolution of the collapsing core. Runs 1 and 2 show an identical evolution until the gas temperature reaches $\sim 35$ K at which point
the two curves fork, with run 2 entering lower temperatures due to a drop in adiabatic index; a light blue trench is clearly visible
in panel (b) between 30 and 80 K while $\gamma_{\text{eff}}$ remains constant in panel (a) all the way up to 80 K. Different treatments
of ortho- and para-hydrogen is therefore the origin of the discrepancies between the thermal evolutions of \citet{tomida2013} and
\citet{vaytet2013}.

In addition, we note that while the variations in dust and gas opacities (destruction of different dust species, sharp rise in atomic
gas opacities at high temperatures) will have an impact on the transport of radiative flux \citep[see][]{vaytet2013}, they do not seem
to affect the thermal evolution of the collapsing system to any great extent.

The thermal evolutions are displayed in more detail in Fig.~\ref{fig:forces}. Panel (e) shows again the density of the gas at the centre
of the grid as a function of its temperature, with a couple of insets revealing more of the curves' features. Inset (i) first unveils the
absence of a bounce\footnote{Period of time during which the core is thermally supported, before collapse resumes when the core has
accreted enough mass \citep[see][]{vaytet2013}} in run 1 compared to runs 2 and 3 (the bounce is labeled \cbone), while on the other hand
inset (j) shows bounce {\cbtwo} which is apparently only present in run 1. Run 1 seems to be sharply forced (by some strange/unknown 
mechanism) to re-join the thermal track of the last two runs for a short time, right before the onset of the second collapse. Panel (g)
displays the evolution of $\gamma_{\text{eff}}$ along the thermal tracks. The difference between the fixed ratio and equilibrium treatment
of ortho- and para-hydrogen is clearly visible for densities in the range $10^{-12} < \rho < 10^{-9}~(\text{g~cm}^{-3})$, and bounces
{\cbone} and {\cbtwo} are again highlighted in insets (k) and (l), respectively.

To determine the origin of these thermal oscillations, we have computed the different forces acting on the fluid and combined them into a
normalised criterion for hydrostatic equilibrium of the first core
\begin{equation}
\Delta F = \frac{F_{\text{pres}} - F_{\text{grav}} - F_{\text{ram}}}{F_{\text{grav}}}
\end{equation}
where
\begin{alignat}{1}
F_{\text{pres}} & = \frac{1}{V} \int_{0}^{R_{s}} \left| \frac{dP_{\text{tot}}}{dr} \right| ~ 4\pi r^{2} dr ~,\\
F_{\text{grav}} & = \frac{1}{V} \int_{0}^{R_{s}} \frac{G M_{\text{enc}} \rho}{r^{2}} ~ 4\pi r^{2} dr ~,\\
F_{\text{ram }} & = \frac{1}{V}  \rho u_{s}^2 ~4\pi R_{s}^{2} ~,
\end{alignat}
and $V = 4/3 \pi R_{s}^{3}$, $R_{s}$, and $u_{s}$ are the first core volume, the accretion shock radius and the gas velocity just
upstream of the accretion shock, respectively. The total pressure $P_{\text{tot}}$ is the sum of the gas pressure $P_{\text{gas}}$ and
the radiative pressure $P_{\text{rad}}$. The force differential $\Delta F$ as a function of central density is shown in
Fig.~\ref{fig:forces}a (the curves are only plotted once the first core has formed, we have assumed this happens at
$t_{\text{0}} = 192~\text{Kyr} \simeq 1.01~t_{\text{ff}}$, when the density at the centre of the grid exceeds
$10^{-13}~\text{g~cm}^{-3}$), with an inset in panel (b) zooming on the detail of the first hydrostatic core. For low densities, the
total pressure force $F_{\text{pres}}$ is clearly over-powered by the gravity ($F_{\text{grav}}$) and ram pressure ($F_{\text{ram}}$)
forces (we consider the ram pressure to be the force applied by the infalling envelope's gas onto the surface of the first core). Then,
as the gas heats up, $P_{\text{tot}}$ rises and brings the system very close to hydrostatic equilibrium. All three runs overshoot the
$\Delta F = 0$ mark, and show subsequent oscillations which die down and disappear for densities above $10^{-8.5}~\text{g~cm}^{-3}$.
Panel (c) also shows the different contributions to $\Delta F$, revealing that the radiative pressure plays a very minor role, while
the ram pressure is important at the first core formation but diminishes once the core has reached quasi-equilibrium.

The overshoots in panels (a) and (b) indicate that all three runs should display a bounce {\cbone} in their thermal tracks; so why is
it not the case for run 1? The temporal evolutions of the central densities and temperatures can provide an answer. Indeed, panels (d)
and (f) show that the clouds in runs 2 and 3 collapse faster than in run 1 (due to the drop in $\gamma_{\text{eff}}$; see
Figs~\ref{fig:rhoT_all}b and \ref{fig:forces}g). As a result, the heating at the centre of the core is very sudden when the value of
$\gamma_{\text{eff}}$ recovers the classical diatomic value of 7/5, causing a sharp oscillation in $\Delta F$, i.e. bounce {\cbone} in
panel (b). Things are a lot smoother for run 1, with a much more steady increase in central density and temperature. Small oscillations
are seen in panels (d) and (f) (as indeed was noticed by \citealt{tomida2013} in their simulations), but nothing large enough to induce
a detectable oscillation in the thermal track. In addition, the overshoot in runs 2 and 3 is actually higher than for run 1.

In the case of bounce {\cbtwo}, what looked like a relatively violent event in panels (e) and (g), and by extension also in inset (m)
of panel (b), turns out to be a relatively slow oscillation which can be seen in panel (f). The small region of the $(\rho,T)$ space
where all three curves overlapped right before the start of the second collapse represents in fact quite a substantial part of the
first core's lifetime. Between 600 and 1100 years after the formation of the first core, there is a lengthy transition period
\citep[see][]{vaytet2013} during which the central density and temperature increase very slowly while the core continues to accrete
mass. Run 1 is not forced to re-join the thermal track of runs 2 and 3 (as was first suggested by panel e), all runs simply eventually
reach the same state of hydrostatic equilibrium where they remain for $\sim500$ years, forget about their early evolution, before they
enter the second phase of the collapse. The short region where all the curves overlap just after bounce {\cbtwo} in inset (j) actually
corresponds to a relatively long period of time in panels (d) and (f).

\begin{figure}
\centering
\includegraphics[scale=0.77]{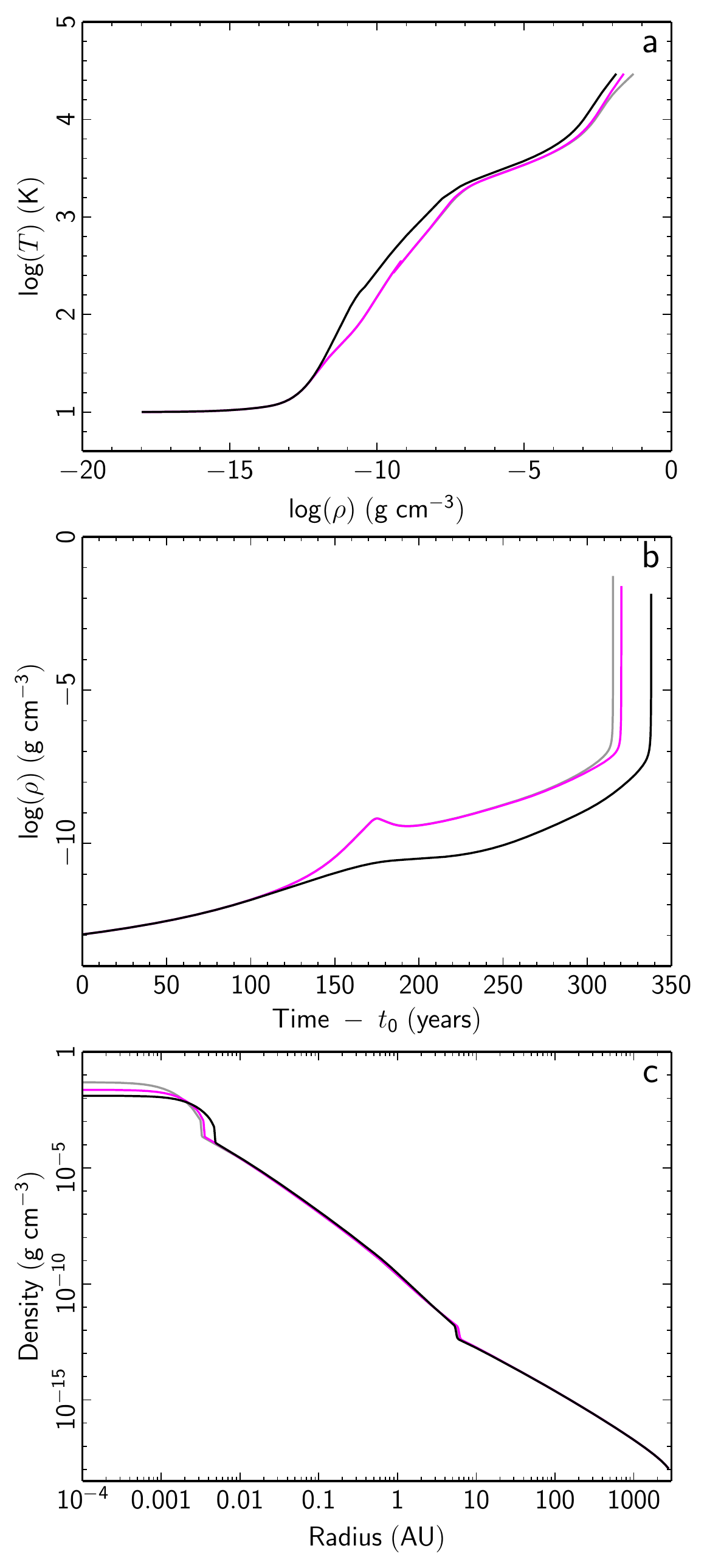}
\caption{Three additional simulations with a parent cloud half the size, using EOS A (black), EOS B (magenta), and EOS C
(grey). (a) Central temperature as a function of central density. (b) Central density as a function of time
($t_{\text{0}} = 62~\text{Kyr} \simeq 0.997~t_{\text{ff}}$). (c) Density radial profiles.}
\label{fig:rhoT_smallbox}
\end{figure}

This was confirmed by running a further three simulations for which we halved the parent cloud size (runs 4, 5, and 6 in
Table~\ref{tab:params}); a smaller cloud radius for the same cloud mass produced thermal evolutions without a transition period in
\citet{vaytet2013}. Figure~\ref{fig:rhoT_smallbox} shows the thermal evolutions for these new simulations (the colour coding for the
different EOSs remains the same). The calculations using an equilibrium OPR still show a bounce around
$\sim 10^{-10} - 10^{-9}~\text{g~cm}^{-3}$ in panel (a) but no flat plateau is visible in panel (b). There is also no lengthy
hydrostatic period in the evolution of the fixed OPR run, and as a consequence the thermal track never meets the tracks from the other
two runs again, it remains at higher temperatures throughout the rest of the simulation. This inevitably has an effect on the second
core properties, which is formed a lower densities and is consequently larger in size. It is also larger in mass, as reported in
Table~\ref{tab:params}.

Finally, during the second phase of the collapse\footnote{The second collapse begins later in run 1 because the $\text{H}_{2}$
dissociation temperature is slightly higher in the fixed 3:1 case than for the equilibrium EOS.}, Fig.~\ref{fig:forces}a shows that the
gravitational force once again overcomes the failing thermal pressure (the thermal energy is being consumed by the dissociation of the
$\text{H}_{2}$ molecules). Hydrostatic equilibrium is recovered at the end of the second collapse, with again some overshoot. The shape
of the curves suggest that more oscillations are to come, but the simulations were stopped before these became apparent (once the second
core is formed, the computational time required to advance the simulations further becomes colossal).

\subsection{Radial profiles}\label{sec:profiles}

\begin{figure*}
\centering
\includegraphics[scale=0.80]{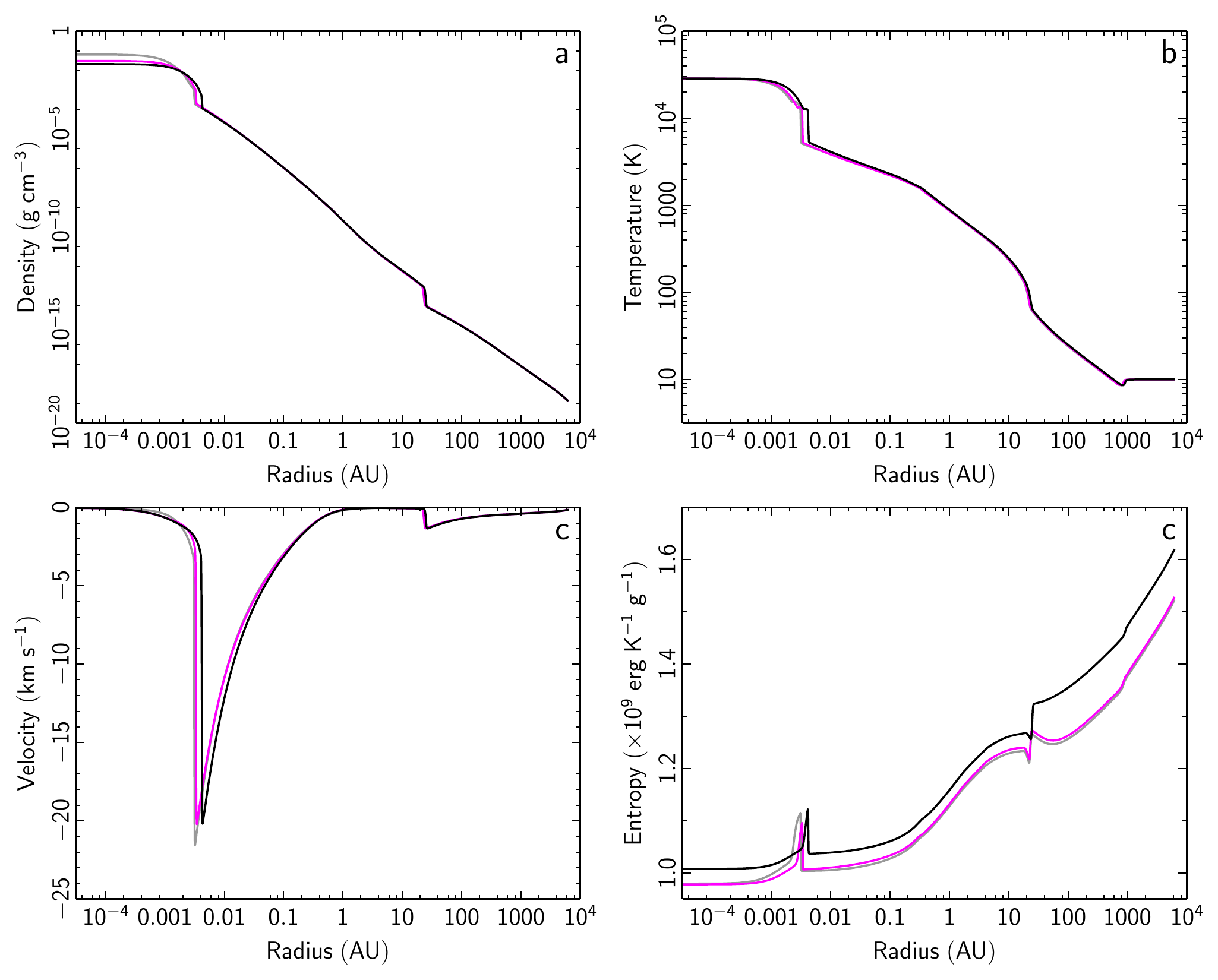}
\caption{Radial profiles of the gas density (a), temperature (b), velocity (c) and entropy (d). As in Fig.~\ref{fig:rhoT_all}, the
black, magenta, and grey curves represent runs 1, 2, and 3, respectively.}
\label{fig:profiles}
\end{figure*}

While the OPR has a significant impact on the thermal evolution of the collapsing cores, it does not seem to affect the radial profiles
of their gas quantities in any great way, as shown in Fig.~\ref{fig:profiles}. The gas density, temperature, and velocity profiles of the
first two runs are extremely similar for most of the system. A small difference ($\sim 25$\%) in second core radius (see
Table~\ref{tab:params}), and an even smaller difference in central density (for the same temperature), are observed. The entropy profiles
differ somewhat more because the partition functions of $\text{H}_{2}$ are different.

In the case of more unstable clouds (runs 4-6), the first cores remain insensitive to the choice of OPR, while some moderately larger
differences arise in the second core's characteristics. As mentioned in Sec~\ref{sec:thermal_evolution}, the thermal tracks of simulations
using EOS A probe higher temperatures compared to runs with EOS B (for the same central density), and the second core consequently forms
at lower densities, with a larger radius. The core is $\sim 40$\% larger in size and mass, and this is clearly visible in
Fig.~\ref{fig:rhoT_smallbox}c (see also Table~\ref{tab:params}). We must however note here that we are comparing second core profiles at
the onset of protostellar formation, and the differences reported here may only be short-lived.

These variations (for both marginally unstable and more unstable clouds) are however probably much too small (and very transient?) to be
detected in observations, and measuring core masses, radii, temperatures or velocity profiles cannot be used to differentiate between a
fixed or an equilibrium ratio of ortho:para hydrogen. Moreover, the disparities are occuring at the second Larson core which is deeply
embedded inside the first code and extremely difficult (if not impossible) to observe directly. The only feasible (albeit difficult)
method to disentangle the two remains the study of chemistry in the system through spectroscopic observations.

At high densities, small differences start to appear between runs 2 and 3. Indeed, Fig.~\ref{fig:rhoT_all}c shows a departure in
$\gamma_{\text{eff}}$ at high densities from ideality; this is a region where plasma effects become important \citep{saumon1995}. This
also results in a slightly higher density for the same temperature (Fig.~\ref{fig:profiles}a). The subsequent evolution of the second
Larson core will probably be affected by non-ideal effects to some extent, but the results of this paper suggest that the ideal mixture
of H and He is a valid description of the state of the gas during the very early stages of star formation.

\subsection{Confusion in other studies}\label{sec:other_studies}

In \citet{vaytet2013}, after comparing several studies of gravitational cloud core collapse, the authors concluded that the OPR was
not the main factor contributing to differences in thermal evolution of the collapsing bodies. The five studies compared were the
works of \citet{masunaga2000,whitehouse2006,stamatellos2007,tomida2013} and \citet{vaytet2013}. The argument was that works which
made use of equilibrium OPR resembled ones assuming non-equilibrium OPR and vice-versa. We now suspect that the OPR is indeed
responsible for the two different thermal tracks, and that the misled conclusions of \citet{vaytet2013} are the result of confusing
terminology employed in the papers.

\citet{masunaga2000} and \citet{vaytet2013} both make use of the \citet{saumon1995} EOS, and it is thus natural that they display
very similar thermal evolutions (a thermal track equivalent to run 3 in the present paper). \citet{tomida2013} uses the non-equilibrium
fixed 3:1 EOS of the present paper, and consequently shows a thermal evolution which strongly resembles run 1 (and run 4 even more).
\citet{whitehouse2006} say that they use the model of \citet{black1975} which claims that equilibrium is assumed at all temperatures,
yet the thermal path taken by their collapsing cloud echoes the non-equilibrium track. As discussed in Sec.~\ref{sec:eos}, we believe
that the heat capacity of the \citet{black1975} EOS is in fact much closer to a non-equilibrium model. Note that they adopt the 3:1 OPR
in their recent works \citep[e.g.][]{bate2011} and the evolution of the non-rotating model is qualitatively consistent with the results
of our non-equilibrium models. Finally, the simulation of \citet{stamatellos2007} follows an equilibrium-like path, even though they
state that their EOS assumes a fixed 3:1 OPR. Interestingly, in a later paper \citep{stamatellos2009}, they use the same numerical
method, reporting that they assume ortho- and para-hydrogen are in equilibrium. It is not clear whether they have changed their
ortho-para strategy between the two studies, or if they simply made a mistake in their \citeyear{stamatellos2007} paper; their thermal
evolution suggests the latter.

We hope this section has cleared up any confusion there may have been between the different studies of gravitational collapse using
radiation hydrodynamics.

\section{Conclusions}\label{sec:conclusions}

We have performed simulations of the gravitational collapse of a dense cloud core using radiation hydrodynamics and distinct gas EOSs
using two different treatments of the OPR; either a fixed 3:1 OPR or an equilibrium ratio. The choice of OPR has a significant impact
on the thermal evolutions of the collapsing cores. Simulations using an equilibrium ratio collapse faster at early times, when a
drop in $\gamma_{\text{eff}}$ (see Figs~\ref{fig:rhoT_all}b and \ref{fig:forces}g) corresponding to increase in $C_{V}$ (see
Fig.~\ref{fig:cv}) facilitates the collapse. This yields rapid heating once $\gamma_{\text{eff}}$ starts to increase again around
$T \sim 100$ K, generating marked oscillations around hydrostatic equilibrium, to the point where the core expands for a short time
right after its formation before resuming its contraction. In the case of a fixed 3:1 OPR, the core's evolution is a lot smoother.
The transition from a monatomic $\gamma_{\text{eff}} = 5/3$ to a diatomic value of 7/5 is monotonous, the thermal support more
important, and hydrostatic equilibrium is reached earlier in terms of central density. By contrast, the radial profiles of the cores
(gas density, velocity, temperature) were not greatly affected by the choice of OPR. First core radii were virtually identical, while
only moderate differences in second core radius and density were observed.

We studied two different initial configurations; marginally unstable and positively unstable parent clouds. In the first case, once
the simulations (both fixed and equilibrium OPR) have reached the first hydrostatic equilibrium, they evolve along the same thermal
tracks, with they central densities and temperatures slowly rising until the second phase of the collapse is triggered by the
dissociation of the $\text{H}_{2}$ molecules, eventually forming very similar second cores. The slow transition period between first
and second collapse was absent from the more unstable cloud calculations, and this implied that the collapsing systems using different
OPRs did not have time to relax towards the same adiabat and consequently did not embark on the second collapse at the same densities.
Fixed OPR simulations yielded an increase of $\sim 40$\% in second core mass and size. We wish to emphasise here that these differences
apply to the newly formed second core, the initial protostellar seed, which will subsequently grow in size and mass at a considerable
rate thanks to the immense accretion rate at the core border. It is very possible that initial difference of 40\% will later become
washed out, once the protostar is well into its evolution towards becoming a young star. We must further acknowledge that our
spherically symmetric simulations do not include any three-dimensional effects such as the launching of outflows and creation of
accretions discs, which play an important role in star formation. 

It is finally not clear which is the best OPR to adopt for simulations of low-mass star formation. While observations suggest that the
OPR is far from equilibrium in dark clouds \citep{pagani2011,dislaire2012}, \citet{flower1984} have shown that under typical molecular
cloud conditions ($n \sim 100 - 1000~\text{cm}^{-3}$) the time when ortho:para equilibrium is reached is of the order of 1 Myr. This
is of course five to ten times larger than the free-fall time of the core we are modelling, but this core is formed as a result of
turbulence in the molecular cloud which spawns over-densities that eventually become gravitationally unstable. The collapse will begin
long after the formation of the molecular cloud which has a typical lifetime of $\sim 10$ Myr and it is thus very possible that at
the onset of the collapse, ortho:para equilibrium has already been reached. In summary, it is not obvious which ortho:para strategy (fixed
ratio or equilibrium) is the most representative of the initial conditions of star formation. Nevertheless, if one is solely interested
in the final properties of the cores when they are formed, it may not matter greatly which OPR is used. On the other hand, if one's
focus lies primarily in the evolution of the first core\footnote{We can only speak for the evolution of the first core since we have
stopped our simulations just after the formation of the second core and did not follow its subsequent evolution.}, the choice of OPR is
of substance. In addition, as \citet{boley2007} pointed out, stability of massive circumstellar disks can also be considerably affected
by the OPR. The typical temperature range found inside protoplanetary discs (apart from the inner disc regions) coincides with the area
where the heat capacities given by the fixed and equilibrium OPR differ the most \citep[$20 - 300$ K; see][for instance]{pinte2009}.
Accretion discs formed using an equilibrium OPR may have a lower temperature and could in principle be more prone to fragmentation, but
this is pure speculation and we will refrain from drawing any conclusions here before having run the simulations in 3D, as many additional
effects (magnetic fields, angular momentum) also come into the picture, possibly weakening the importance of the OPR.

\begin{acknowledgements}
The research leading to these results has received funding from the European Research Council under the European Community's Seventh 
Framework Programme (FP7/2007-2013 Grant Agreement no. 247060). KT is supported by Japan Society for the Promotion of Science (JSPS)
Postdoctoral Fellowship for Research Abroad.
\end{acknowledgements}

\end{document}